\documentclass{aa}  
\usepackage{graphicx}
\usepackage{txfonts}
\usepackage{natbib}
\bibpunct{(}{)}{;}{a}{}{,} 

\newcommand{\gd}{GD~99}

\usepackage[normalem]{ulem}

\begin{document}

   \title{\gd: Re-investigation of an old ZZ~Ceti companion}


    \author{Zs.~Bogn\'ar\inst{1,2,3}\fnmsep\thanks{\email{bognar.zsofia@csfk.org}},
        \'A.~S\'odor\inst{1,2,3},
        \and Gy.~Mez\H o\inst{1,3}}
   
   \institute{
        Konkoly Observatory, E\"otv\"os Lor\'and Research Network (ELKH), Research Centre for Astronomy and Earth Sciences, Konkoly Thege Mikl\'os \'ut 15-17, H--1121, Budapest, Hungary
        \and
        MTA CSFK Lend\"ulet Near-Field Cosmology Research Group
        \and
        MTA Centre of Excellence
        }
        
    \titlerunning{\gd\ re-investigated}
        \authorrunning{Zs.~Bogn\'ar et al.}
        
    \date{}

 
  \abstract
   {Thanks to photometric space missions, we have access to more and more information on the properties of white dwarf stars, especially pulsating ones.
   In the case of pulsators, we have the opportunity to get an insight into their otherwise hidden interiors by the means of asteroseismology. In addition to space-based observations, we also take advantage of the opportunity to study the pulsations of white dwarf stars from the ground, either as observations that are complementary to space-based measurements or individual observing runs on selected targets across long timescales.}
   {We aim to investigate long-term, single-site observations of the bright, yet scarcely studied ZZ~Ceti star, \gd. Our main goals are to determine as many eigenmodes for asteroseismology as possible and then to carry out a seismic analysis of this target.}
   {We performed a Fourier analysis of the light curves obtained in different epochs. After finding the normal modes of the pulsation, we ran the 2018 version of the White Dwarf Evolution Code to build model grids for the period fits. We compared the seismic distance of the best-fit model with the geometric value provided by Gaia measurements.}
   {We find that \gd\ is rich in pulsation modes in the $\sim200-1100\,$s period range, as we detected seven new periods. Together with the literature data, we were able to use 11 modes for the asteroseismic fits. We accepted an asteroseismic model solution with $T_{\mathrm{eff}}=13\,500\,$K and $M_*=0.80\,M_{\odot}$ as a best fit, however, this suggests a hotter and more massive star than we might have expected based on the spectroscopic values. We also estimated the rotational rate of the star to be $13.17\,$h, based on TESS observations.}
   {}


   \maketitle
%

\section{Introduction}

The star \gd\ is one of the first  ZZ~Ceti variables to be discovered. Its light variations were observed by \citet{1976ApJ...205L.155M} for the very first time. The pulsation periods were published by \citet{2006ApJ...640..956M}.
Apart from this, no additional photometric observations have been published so far, despite the fact that \gd\ is one of the brightest known ZZ~Ceti stars. Thus, we selected this target for further, long-term and ground-based photometric observations, as one of the targets in our observing programme aimed at obtaining measurements as well as (in many cases) carrying out  an asteroseismological investigation of our target objects (see e.g. \citealt{2009MNRAS.399.1954B, 2013MNRAS.432..598P, 2014A&A...570A.116B, 2016MNRAS.461.4059B, 2018MNRAS.478.2676B, 2019MNRAS.482.4018B, 2021A&A...651A..14B, 2021AcA....71..281K}).

ZZ~Ceti (a.k.a. DAV stars) represent the most populous group amongst white dwarf (WD) variables. Most of these stars (about $97\%$) end their evolution as WDs and occupy a wide range of effective temperatures from the extremely hot (about 200\,000\,K) to the cooled-down objects around 4000\,K. They can exhibit a broad mass range,  from $0.15$ to $1.3$ solar masses, although most of them are between $0.5 - 0.7\,M_\sun$. As these stellar remnants are about Earth-sized, their surface gravities are relatively high, namely, log\,$g \sim 8$, and because of the gravitational settling, we find the heavier constituting elements at deeper levels than the lighter ones. Considering a representative DA spectral type white dwarf, it has a degenerate carbon-oxygen (C-O) core, surrounded by a non-degenerate thin helium (He) layer and an even thinner hydrogen (H) one.

As these stars cool and reach the so-called ZZ~Ceti instability strip around $10\,500 - 13\,000\,$K, they become pulsationally unstable, displaying low-amplitude ($\sim$ mmag), short-period ($100-1500\,$s) $g$-mode pulsations, which we can detect as brightness variations that characterise the stars due to surface temperature changes. The excitation is provided by the classical $\kappa-\gamma$ mechanism \citep{1981A&A...102..375D, 1982ApJ...252L..65W} in combination with convective driving \citep{1991MNRAS.251..673B, 1999ApJ...511..904G}. We would have thought that the ZZ~Ceti stars behave similarly through the entire DAV instability strip, however, the observations contradict this presumption. As clearly described, for instance, in \citet{2017ApJS..232...23H}, the pulsational behaviour of the WDs depends on their situation along the instability strip: closer to the hot (blue) edge, the stars show the lowest amplitude and shortest period light variations; in the middle of the strip, we observe much higher amplitudes; and as we are approaching the cool (red) edge of the instability region, the light-variation amplitudes start to decrease with longer and longer periods. We also observe the changing stability of the pulsational behaviour with changing temperature. Cooler variables are more likely to show short-term (from days to weeks) amplitude and frequency variations, while the pulsation properties are more stable of the hotter objects, close to the blue edge of the instability strip.
Space observations have revealed that some of the  cool ZZ~Ceti stars demonstrate irregularly recurring outburst events, when the stellar flux can increase up to 15\% \citep{2017ASPC..509..303B}.
For comprehensive reviews on the characteristics of pulsating white dwarf stars, see the papers of \citet{2008ARA&A..46..157W}, \citet{2008PASP..120.1043F}, \citet{2010A&ARv..18..471A}, \citet{2019A&ARv..27....7C}, and \citet{2020FrASS...7...47C}.

Pulsating white dwarfs are otherwise recognised as regular white dwarf stars. These are ideal space laboratories for studying the behaviour of matter under extreme physical conditions. However, the only way we can investigate the interiors of these stars is through the study of the global oscillations excited in them, via the apparatus of  asteroseismology. Our primary goal is to determine as many as possible pulsational frequencies in oscillating stars in order to impose stronger constraints for asteroseismic modelling.

We present our ground-based observations from different years, along with measurements obtained by the Transiting Exoplanet Survey Satellite (TESS; \citealt{2015JATIS...1a4003R}) in Sect.~\ref{sect:obs}, followed by the light curve analyses of the data sets in Sects.~\ref{sect:analg} and \ref{sect:analt}. We summarise our period findings in Sect.~\ref{sect:periods}. The asteroseismic analysis is presented in Sect.~\ref{sect:model}. The summary and discussion of our results are given in Sect.~\ref{sect:sum}.

\section{Observations and data reduction}
\label{sect:obs}

We observed \gd\ ($G=14.56$\,mag, $\alpha_{2000}=09^{\mathrm h}01^{\mathrm m}49^{\mathrm s}$, $\delta_{2000}=+36^{\mathrm d}07^{\mathrm m}08^{\mathrm s}$) on 30 nights from the ground and we also obtained data with the TESS space telescope.

\subsection{Ground-based observations}

We performed our observations between 2002 and 2022 with the 1m Ritchey--Chr\'etien--Coud\'e telescope at the Piszk\'estet\H o mountain station of Konkoly Observatory, Hungary. The detector was equipped with a Photometrics Thomson TH7896M CCD chip and we applied 40\,s exposure times in addition to $\sim20$\,s readout times in 2002. The following observations were performed in 2018 with an FLI Proline 16803 CCD camera and with a read-out time of $\sim3$\,s. We continued the observations in the 2020/2021 observing season, when the telescope was equipped with the Spectral Instruments (SI) 1100S CCD. The readout time of this camera is $\sim22$\,s. In the 2021-2022 observing season we utilised two cameras: SI 1100S CCD and Andor iXon+888 Electron-multiplying CCD (EMCCD). In the latter case the readout time was $\sim4$\,s. In all cases (except in 2002), we observed in white light to maximise the photon counts, whereas in 2002, we observed in  the $V$ band. The observing log is presented in Table~\ref{tabl:log}, with the corresponding dates, exposure times, numbers of data points, and observing lengths included.

\begin{table}
\centering
\caption{Log of ground-based observations of \gd. `Exp' is the integration time, \textit{N} is the number of data points, and $\delta T$ is the length of the data sets including gaps. Weekly observations in the case of the SI camera are denoted by `a--g' letters in parentheses. We also list the cameras used for the corresponding observations.}
\label{tabl:log}
\tiny
\begin{tabular}{lrccrr}
\hline
\hline
Run & UT Date & Start time & Exp. & \textit{N} & $\delta T$ \\
 &  & (BJD-2\,450\,000) & (s) &  & (h) \\
\hline
\multicolumn{6}{l}{Photometrics Inc. CCD} \\
01 & 2002 Feb 04 & 2310.274 & 40 & 519 & 9.46 \\
02 & 2002 Feb 05 & 2311.276 & 40 & 542 & 9.53 \\
\multicolumn{2}{l}{Total:} & & \multicolumn{2}{r}{1061} & 18.99\\
\multicolumn{6}{l}{FLI Proline CCD} \\
03 & 2018 Oct 12 & 8403.540 & 30 & 309 & 2.87 \\
04 & 2018 Oct 14 & 8405.548 & 30 & 284 & 2.64 \\
05 & 2018 Oct 15 & 8406.545 & 30 & 284 & 2.69 \\
\multicolumn{2}{l}{Total:} & & \multicolumn{2}{r}{877} & 8.20\\
\multicolumn{6}{l}{SI CCD} \\
06(a) & 2021 Feb 15 & 9261.248 & 30 & 426 & 6.27 \\
07(b) & 2021 Mar 15 & 9289.320 & 30 & 409 & 6.01 \\
08(b) & 2021 Mar 16 & 9290.308 & 30 & 450 & 6.80 \\
\multicolumn{2}{l}{Total:} & & \multicolumn{2}{r}{1285} & 19.08\\
\multicolumn{6}{l}{EMCCD} \\
09 & 2021 Nov 05 & 9523.505 & 30 & 170 & 1.61 \\
10 & 2021 Nov 05 & 9524.448 & 30 & 601 & 5.68 \\
11 & 2021 Nov 08 & 9527.478 & 30 & 500 & 4.73 \\
12 & 2021 Nov 09 & 9528.463 & 30 & 524 & 4.97 \\
\multicolumn{2}{l}{Total:} & & \multicolumn{2}{r}{1795} & 16.99\\
\multicolumn{6}{l}{SI CCD} \\
13(c) & 2021 Dec 07 & 9556.400 & 30 & 467 & 7.55 \\
14(d) & 2022 Jan 06 & 9586.308 & 30 & 664 & 9.97 \\
15(d) & 2022 Jan 07 & 9587.320 & 30 & 221 & 3.20 \\
16(d) & 2022 Jan 10 & 9589.527 & 30 & 329 & 4.91 \\
17(d) & 2022 Jan 10 & 9590.437 & 30 & 446 & 6.80 \\
18(d) & 2022 Jan 11 & 9591.311 & 30 & 658 & 9.71 \\
19(d) & 2022 Jan 12 & 9592.317 & 30 & 589 & 8.64 \\
20(e) & 2022 Feb 11 & 9622.430 & 30 & 402 & 6.15 \\
21(e) & 2022 Feb 12 & 9623.229 & 30 & 711 & 11.06 \\
22(f) & 2022 Feb 24 & 9635.239 & 30 & 262 & 4.95 \\
23(f) & 2022 Feb 27 & 9638.252 & 30 & 247 & 3.58 \\
24(f) & 2022 Feb 28 & 9639.247 & 30 & 452 & 7.21 \\
25(f) & 2022 Mar 01 & 9640.236 & 30 & 455 & 7.72 \\
26(g) & 2022 Mar 24 & 9663.267 & 30 & 261 & 3.81 \\
27(g) & 2022 Mar 25 & 9664.326 & 30 & 308 & 4.93 \\
28(g) & 2022 Mar 26 & 9665.303 & 30 & 448 & 6.76 \\
29(g) & 2022 Mar 27 & 9666.262 & 30 & 398 & 5.85 \\
30(g) & 2022 Mar 28 & 9667.261 & 30 & 395 & 6.54 \\
\multicolumn{2}{l}{Total:} & & \multicolumn{2}{r}{7713} & 119.34\\
\hline
\end{tabular}
\end{table}

\begin{figure*}
\centering
\includegraphics[width=0.9\textwidth]{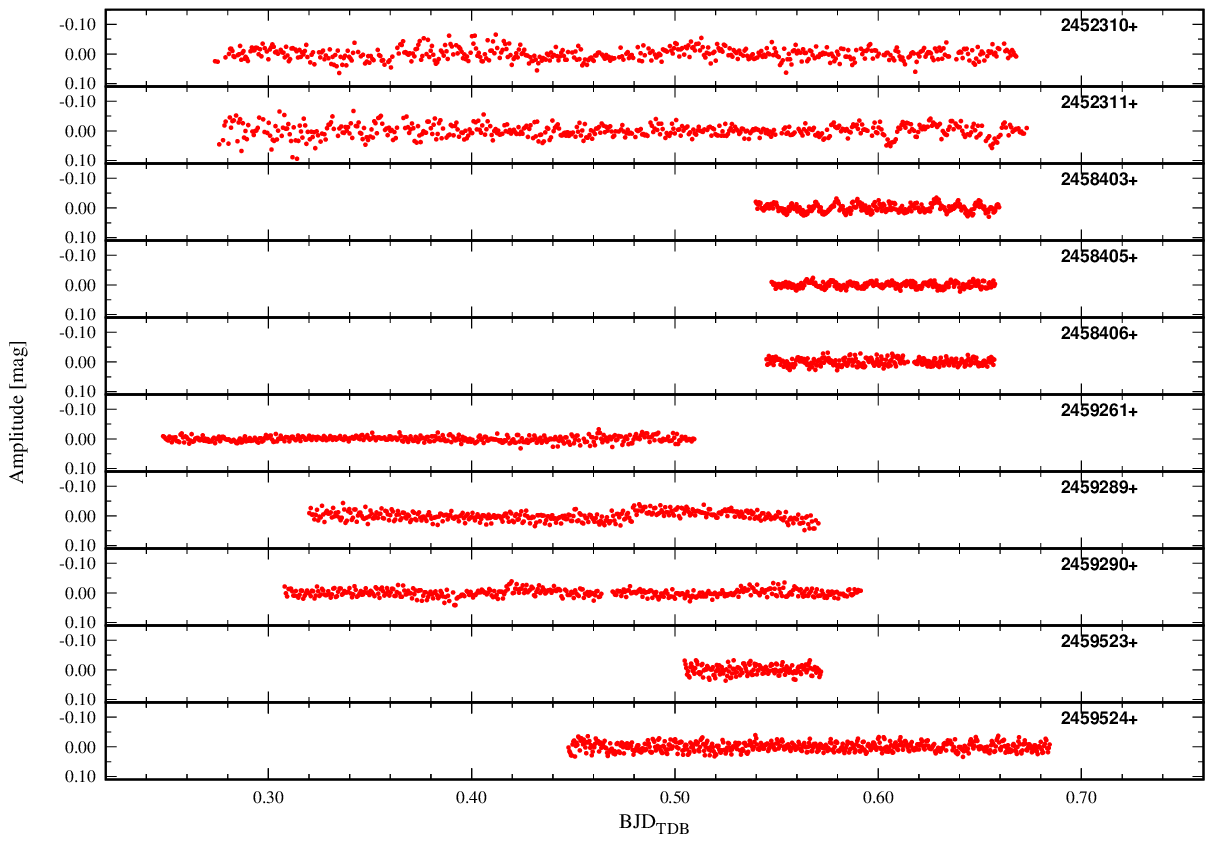}
\includegraphics[width=0.9\textwidth]{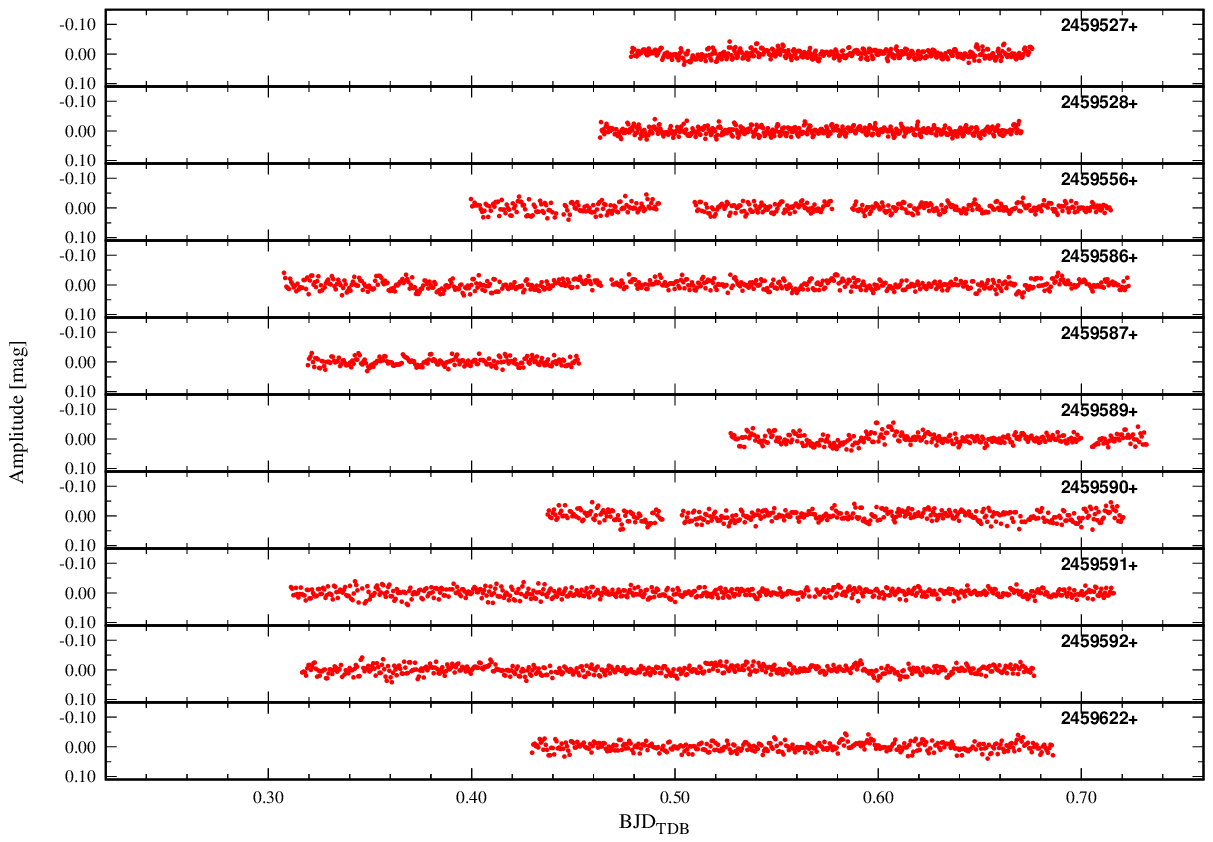}
\caption{Normalised differential light curves of \gd\ -- part one.}{\label{fig:lc_all1}}
\end{figure*}

\begin{figure*}
\centering
\includegraphics[width=0.9\textwidth]{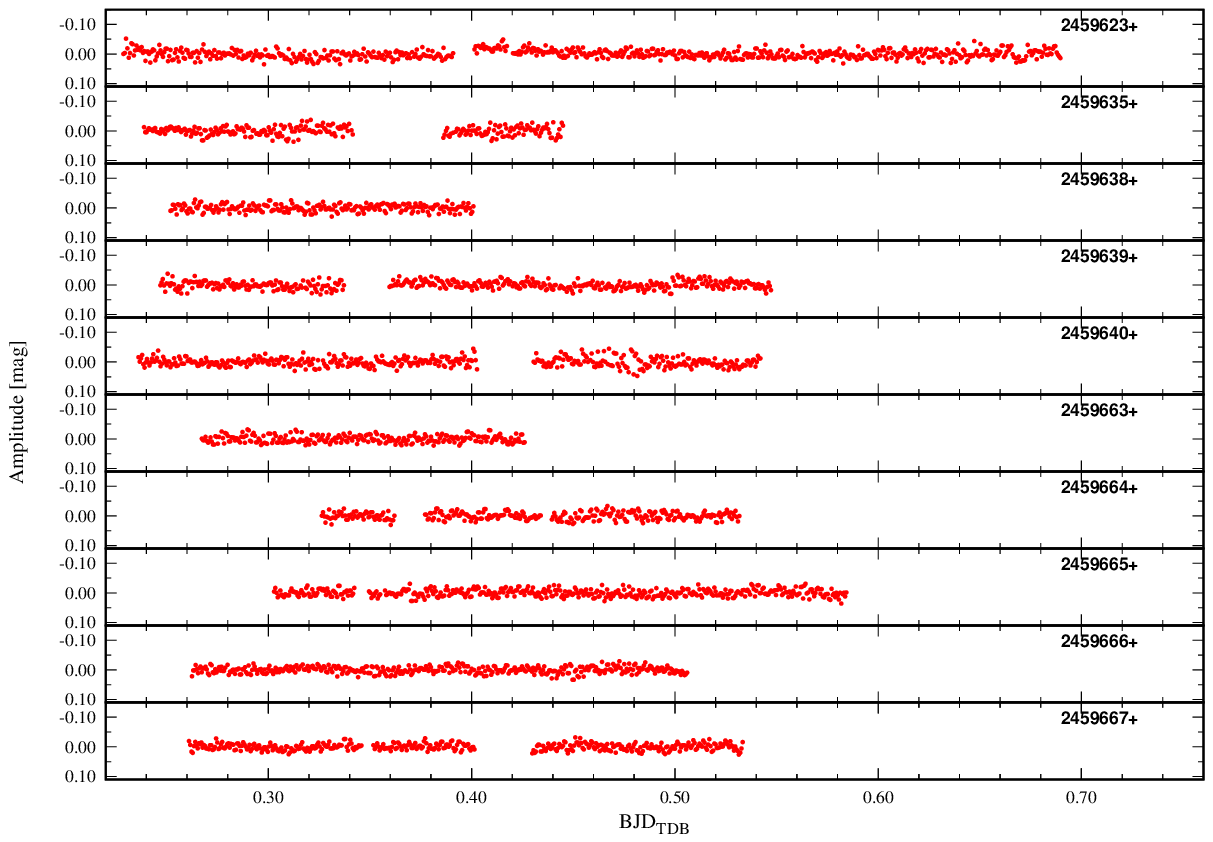}
\caption{Normalised differential light curves of \gd\ -- continued.}{\label{fig:lc_all2}}
\end{figure*}

We performed the reduction of the raw data frames the standard way utilising \textsc{iraf}\footnote{\textsc{iraf} is distributed by the National Optical Astronomy Observatories, which are operated by the Association of Universities for Research in Astronomy, Inc., under a cooperative agreement with the National Science Foundation.} tasks. After bias, dark, and flat corrections, we carried out the aperture photometry process for the field stars. We fit second- or third-order polynomials to the resulting light curves, correcting for long-period atmospheric and instrumental trends. We converted the observational times of every data point to barycentric Julian dates in barycentric dynamical time (BJD$_\mathrm{{TDB}}$) using the applet of \citet{2010PASP..122..935E}\footnote{http://astroutils.astronomy.ohio-state.edu/time/utc2bjd.html}.
The panels of Figs.~\ref{fig:lc_all1} and \ref{fig:lc_all2} show the plots of all ground-based data obtained in chronological order. 

\subsection{TESS observations}

\begin{figure}
\centering
\includegraphics[width=0.45\textwidth]{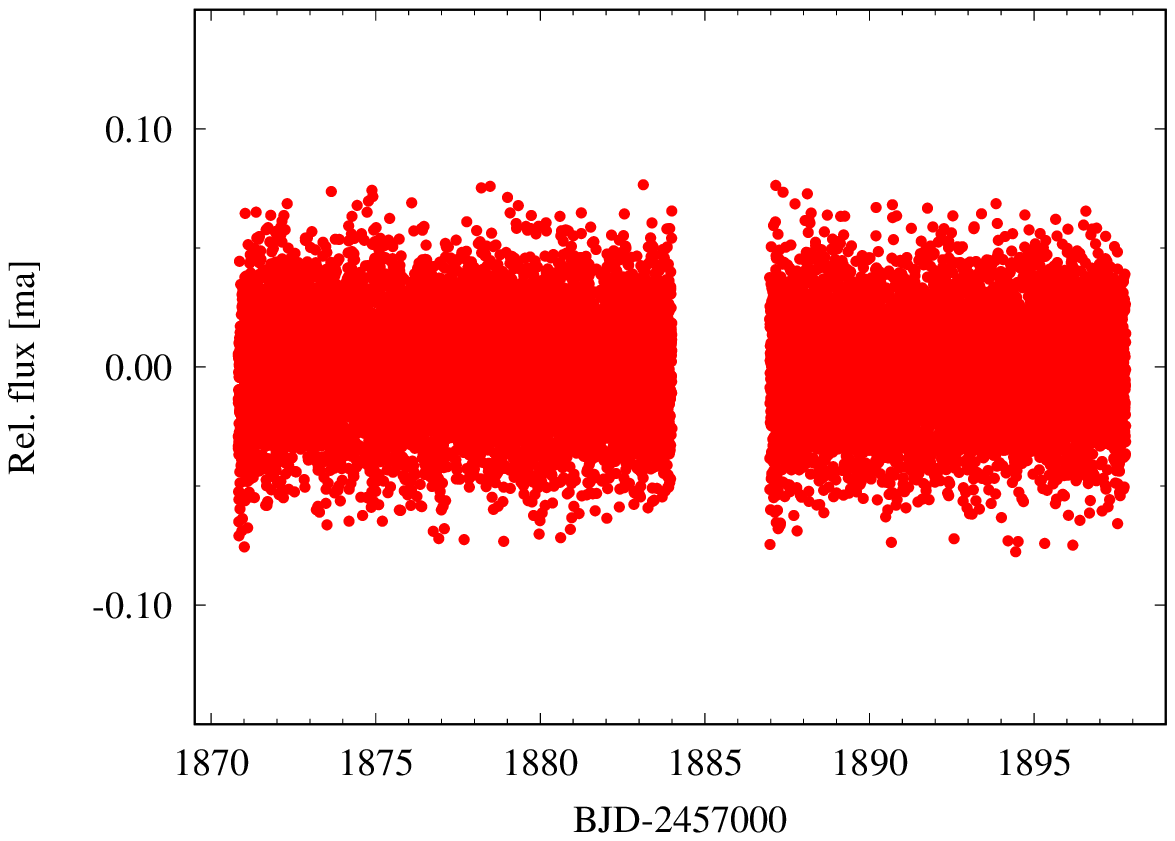}
\caption{Light curve of \gd\ obtained by TESS.}{\label{fig:tesslc}}
\end{figure}

TESS observed \gd\ for $27.4$ days in sector 21 with the 120 second short-cadence mode between 21 January and 18 February 2020. We downloaded the light curves from the \textit{Mikulski Archive for Space Telescopes} (MAST), and extracted the PDCSAP fluxes provided by the Pre-search Data Conditioning Pipeline \citep{2016SPIE.9913E..3EJ}.  We omitted the obvious outliers. The resulting light curve consists of 17\,005 data points (with a gap). The cleaned light curve can be seen in Fig~\ref{fig:tesslc}.

\section{Light curve analyses of the ground-based observations}
\label{sect:analg}

For the standard Fourier analysis of the data sets, we utilised the photometry modules of the Frequency Analysis and Mode Identification for Asteroseismology (\textsc{famias}) software package \citep{2008CoAst.155...17Z}. Considering the experience we gained investigating the light curves of several other white dwarf variables, we accepted a frequency peak as significant if its amplitude reached the  signal-to-noise ratio (S/N) limit of 5 instead of the widely used 4~S/N limit. We calculated the noise level by the average Fourier amplitude in a $\sim 1700\,\mu$Hz radius vicinity ($150\,$d$^{-1}$) of the peak in question (see e.g. in \citealt{2019MNRAS.482.4018B}). We chose the 5~S/N significance level, as below this, we can find several peaks with similarly low amplitudes in many cases, which may often be the result of the amplitude-frequency-phase variations that are frequently detected in white dwarf variables.

We analysed the data sets obtained with different CCDs separately. In the case of the Photometrics, FLI Proline, and Andor EMCCD, we observed \gd\ on one observing week, respectively, whereas we observed the star on seven weeks altogether (weeks a--g in Table~\ref{tabl:log}) with the SI CCD camera. Besides the simple analysis of the weekly data, we tested our frequency solutions by analysing combinations of different consecutive weekly observations. These subsets were selected from the SI-observations and consist of data taken in the weeks (a+b+c), (b+c+d), (c+d+e), (d+e+f), and (e+f+g). For details, we refer to the markings in Table~\ref{tabl:log}.

We identified closely spaced peaks in the $935 - 1045\,\mu$Hz frequency range, while we see more or less well-separated peaks at around $1150$, $1685$, $2890$, $3340$, $3500$, $4390$, and $4525\,\mu$Hz. We reviewed the frequency analysis results of the weekly, as well as the combined weekly data, and we searched for similar independent frequencies at the different frequency domains that we could accept as real pulsation frequencies. We summarise our findings in Table~\ref{tabl:freq1}.

We identified 11 frequencies (listed in Table~\ref{tabl:freq1}), however, the presence of further frequencies is possible at $1030$, $1175$, and $2840\,\mu$Hz, but we do not list them in Table~\ref{tabl:freq1}. At the $\sim1030\,\mu$Hz frequency peak, there are $1\,$d$^{-1}$ alias ambiguities and in the other two cases, we can find a significant peak only in one data set, whereas we were looking for repeatedly appearing frequencies. The only exception we see (in Table~\ref{tabl:freq1}) is the peak at $1147.49\,\mu$Hz. We accepted this frequency as it was the dominant peak in the 2018 (FLI) data set. We note that week (b), which consists of two nights only, is not listed in Table~\ref{tabl:freq1}, because the only significant peak found in this data is at $4374\,\mu$Hz, which is about $-1$\,d$^{-1}$ distance from the $4386\,\mu$Hz peak, is present in almost all of the other data sets. We also note that in many cases, we can detect a peak at $4374\,\mu$Hz or at the $4374 - 1\,$d$^{-1}\,\mu$Hz frequency. This suggests that there is another, closely spaced frequency to the $4386\,\mu$Hz peak. We resolved this ambiguity by analysing the TESS data (see Sect.~\ref{sect:analt}).
We also analysed the complete data set of the most extensively observed season, SI 2021/2022 (18 nights). These results are also listed in Table~\ref{tabl:freq1}.

\begin{table*}
\centering
\caption{Collection of similar independent frequencies at different frequency domains detected by the data sets of \gd\ obtained at different epochs and in the combined weekly data subsets.}
\label{tabl:freq1}
\tiny
\begin{tabular}{lrrrrrrrrrrr}
\hline
\hline
 & \multicolumn{11}{c}{frequency [$\mu$Hz]} \\
\hline
Photometrics Inc. CCD & -- & 945.46 & -- & -- & -- & -- & -- & -- & -- & -- & -- \\
FLI & -- & -- & -- & -- & 1147.49 & -- & -- & -- & -- & 4396.15 & -- \\
EMCCD & -- & 948.52 & -- & 1039.09 & -- & -- & -- & 3337.35 & 3495.87 & 4386.02 & 4515.30 \\
SI week(a) & -- & -- & -- & -- & -- & -- & -- & -- & -- & 4388.81 & -- \\
SI week(c) & 939.68 & -- & 983.81 & -- & -- & -- & -- & -- & 3492.07 & 4385.78 & -- \\
SI week(d) & 934.89 & 943.34 & -- & 1041.39 & -- & -- & -- & -- & 3496.37 & 4385.97 & 4524.27 \\
SI week(e) & -- & 945.31 & -- & -- & -- & -- & -- & -- & -- & 4385.86 & -- \\
SI week(f) & 934.20 & -- & -- & -- & -- & -- & -- & -- & -- & 4386.09 & -- \\
SI week(g) & -- & -- & -- & -- & -- & 1693.77 & 2890.51 & -- & 3486.03 & 4386.24 & -- \\
SI week(a+b+c) & -- & 939.11 & -- & -- & -- & -- & -- & -- & -- & 4383.04 & -- \\
SI week(b+c+d) & 935.03 & -- & 993.03 & 1043.95 & -- & -- & -- & 3335.57 & 3496.41 & 4386.00 & 4524.21 \\
SI week(c+d+e) & 935.09 & 943.49 & 993.06 & 1044.01 & -- & -- & -- & -- & 3496.44 & 4386.04 & 4524.18 \\
SI week(d+e+f) & 934.68 & 943.14 & -- & 1041.19 & -- & -- & -- & -- & 3496.13 & 4386.08 & 4524.56 \\
SI week(e+f+g) & 933.86 & -- & -- & -- & -- & 1685.64 & 2890.16 & 3337.53 & 3497.65 & 4386.15 & 4526.15 \\
 & & & & & & & & & & & \\
SI (2021/2022) & 934.69 & 943.13 & 993.05 & 1044.00     & -- & 1685.61 & 2890.16 & 3337.53 & 3496.82 & 4386.08 & 4526.70 \\
\hline
\end{tabular}
\end{table*}

\section{Light curve analysis of the TESS observations}
\label{sect:analt}

We also performed a frequency analysis on the TESS data. Surprisingly, the only two frequencies above the 5~S/N significance level are at $3948.04$, and $3959.86\,\mu$Hz, with $\sim 1\,$d$^{-1}$ separation. We cannot see peaks at these frequencies in the ground-based data, so we can assume that we have found additional frequencies to those listed in Table~\ref{tabl:freq1}. However, as we investigate the Fourier transform (FT) of the TESS data set above the Nyquist limit ($4167\,\mu$Hz), we can find the pairs of these frequencies at the already known $4385.38$ and $4373.58\,\mu$Hz; that is, the $3948.04$, and $3959.86\,\mu$Hz peaks are Nyquist aliases of the real frequencies at $4385$ and $4374\,\mu$Hz. The TESS data revealed that there is a doublet at this frequency domain, which is also suggested by the ground-based data. The separation of the frequency components is very close to $1\,$d$^{-1}$, which explains why we could not unambiguously determine the lower-amplitude $4374\,\mu$Hz component in the ground-based data. 

\section{Periods for asteroseismology}
\label{sect:periods}

Our main goal with these several weeks of observations is to determine periods as inputs for asteroseismology. We decided to utilise the SI 2021/2022 data set, consisting of 18 nights, as a reference for seismology with ten observed periods. We complemented this list with the dominant frequency of the FLI (2018) data set. This means 11 periods for the asteroseismic investigations. Table~\ref{tabl:final} lists the corresponding frequencies, periods, and amplitudes. As we see, the star shows both long- and short-period light variations. However, we must keep in mind that observations performed at different epochs can reveal additional pulsation modes for a given star. For this reason, we complemented our list of pulsational modes with the periods presented by \citet{2006ApJ...640..956M} for our asteroseismic investigations. We compare our observational results with those presented in \citet{2006ApJ...640..956M} in Table~\ref{tabl:comp}. In the case of closely spaced periods, which can be treated as common frequencies, we calculated the amplitude weighted mean periods for asteroseismology. The final list of periods we used for asteroseismic fits can be found in the last column of Table~\ref{tabl:comp}. As we see, \gd\ is rich in pulsation frequencies, with 18 periods listed.  

\begin{table}
\centering
\caption{List of the set of accepted frequencies by the ground-based observations presented in this work.}
\label{tabl:final}
\begin{tabular}{lrrrr}
\hline
\hline
 & \multicolumn{1}{c}{\textit{f}} & \multicolumn{1}{c}{\textit{P}} & Ampl. \\
 & \multicolumn{1}{c}{[$\mu$Hz]} & \multicolumn{1}{c}{[s]} & [mmag] \\
\hline
$f_{01}$ & 4386.10 & 227.99 & 7.40\\
$f_{02}$ & 1147.49 & 871.47 & 6.38\\
$f_{03}$ & 934.69 & 1069.88 & 2.75\\
$f_{04}$ & 3496.82 & 285.97 & 2.10\\
$f_{05}$ & 943.12 & 1060.31 & 1.95\\
$f_{06}$ & 4526.69 & 220.91 & 1.63\\
$f_{07}$ & 3337.54 & 299.62 & 1.41\\
$f_{08}$ & 993.05 & 1007.00 & 1.36\\
$f_{09}$ & 1044.00 & 957.85 & 1.23\\
$f_{10}$ & 2890.16 & 346.00 & 1.18\\
$f_{11}$ & 1685.61 & 593.26 & 1.16\\
\hline
\end{tabular}
\end{table}

\begin{table*}
\centering
\caption{Comparison of the periods presented this work with the solution listed in \citet{2006ApJ...640..956M}. The last column shows the periods utilised for the asteroseismic fits.}
\label{tabl:comp}
\begin{tabular}{lrrrrr}
\hline
\hline
 & \multicolumn{2}{c}{This work} & \multicolumn{2}{c}{\citet{2006ApJ...640..956M}} & \multicolumn{1}{c}{For seismology} \\
 & \multicolumn{1}{c}{\textit{P}} & Ampl. & \multicolumn{1}{c}{\textit{P}} & Ampl. & \multicolumn{1}{c}{\textit{P}} \\
 & \multicolumn{1}{c}{[s]} & [mmag] & \multicolumn{1}{c}{[s]} & [mma] & \multicolumn{1}{c}{[s]} \\
\hline
 & -- & -- & 105.2 & 2.0 & 105.2 \\
$f_{06}$ & 220.91 & 1.63 & 223.6 & 2.9 & 222.6 \\
$f_{01}$ & 227.99 & 7.40 & 228.9 & 4.5 & 228.3 \\
$f_{04}$ & 285.97 & 2.10 & -- & -- & 286.0 \\
$f_{07}$ & 299.62 & 1.41 & -- & -- & 299.6 \\
$f_{10}$ & 346.00 & 1.18 & -- & -- & 346.0 \\
$f_{11}$ & 593.26 & 1.16 & -- & -- & 593.3 \\
 & -- & -- & 633.1 & 2.0 & 633.1 \\
 & -- & -- & 853.2 & 2.4 & 853.2 \\
$f_{02}$ & 871.47 & 6.38 & -- & -- & 871.5 \\
 & -- & -- & 924.7 & 1.7 & 924.7 \\
$f_{09}$ & 957.85 & 1.23 & -- & -- & 957.9 \\
 & -- & -- & 976.0 & 2.1 & 976.0 \\
$f_{08}$ & 1007.00 & 1.36 & 1007.0 & 6.5 & 1007.0 \\
$f_{05}$ & 1060.31 & 1.95 & 1058.0 & 8.3 & 1058.4 \\
$f_{03}$ & 1069.88 & 2.75 & -- & -- & 1069.9 \\
 & -- & -- & 1088.0 & 4.3 & 1088.0 \\
 & -- & -- & 1151.0 & 1.9 & 1151.0 \\
\hline
\end{tabular}
\end{table*}

\begin{figure*}
\centering
\includegraphics[width=1.0\textwidth]{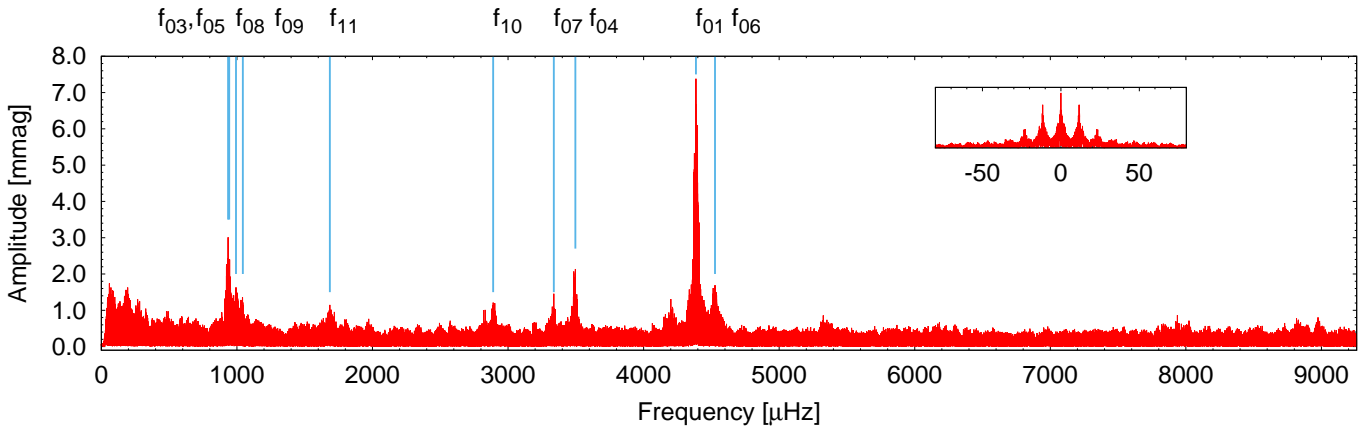}
\caption{Fourier transform of the SI 2021/2022 data set. We also marked the frequencies listed in Table~\ref{tabl:final} for completeness.}{\label{fig:SI}}
\end{figure*}

\begin{figure}
\centering
\includegraphics[width=0.45\textwidth]{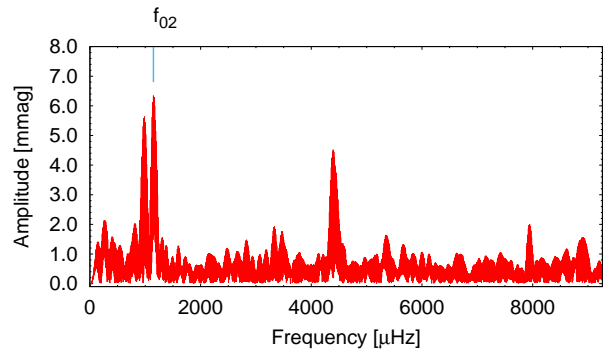}
\caption{Fourier transform of the FLI (2018) data set. We marked the $f_{02}$ frequency listed in Table~\ref{tabl:final}.}{\label{fig:FLI}}
\end{figure}

\begin{figure}
\centering
\includegraphics[width=0.45\textwidth]{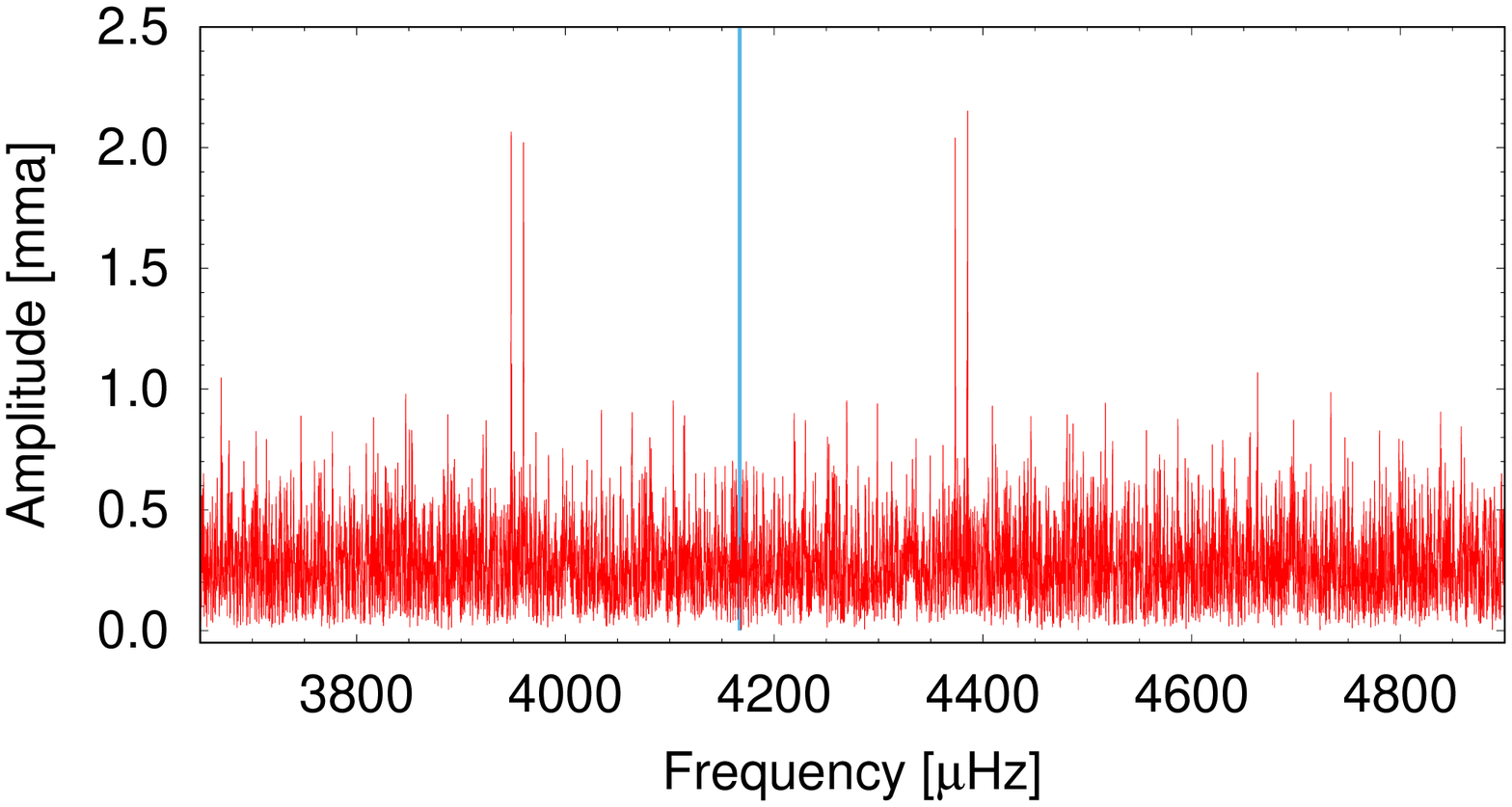}
\caption{Part of the Fourier transform of the TESS data. The vertical blue line denotes the Nyquist limit of the 120\,s cadence observations.}{\label{fig:TESSft}}
\end{figure}

Figures~\ref{fig:SI} and \ref{fig:FLI} show the FTs of the SI 2021/2022 and FLI (2018) data sets, respectively. We marked the accepted frequencies listed in Table~\ref{tabl:final} in these plots. 
The FT of the TESS data can be seen in Fig.~\ref{fig:TESSft}. The doublet at $3948.04$ and $3959.86\,\mu$Hz appears clearly, together with their super-Nyquist pairs at $4385.38$ and $4373.58\,\mu$Hz.

\section{Asteroseismology}
\label{sect:model}

A new version of the White Dwarf Evolution Code (\textsc{wdec}) was presented in 2018 by \citet{2018AJ....155..187B}. It now uses the Modules for Experiments In Stellar Astrophysics (\textsc{mesa}) (\citealt{2011ApJS..192....3P}; version r8118) opacity routines and equation of states. We utilised this \textsc{wdec} version to build model grids for the asteroseismic investigations of \gd. 
We started every model with a $\sim100\,000$\,K polytrope, which was then evolved down to the requested temperature. Finally, we obtain a thermally relaxed solution to the stellar structure equations. The programme treats the convection operates in the white dwarf envelopes within the framework of the mixing length theory \citep{1971A&A....12...21B}. The code permits us to define how to treat the $\alpha$ parametrization. We chose to take $\alpha$ according to the results of \citet{2015ApJ...799..142T}.

We determined the $\ell=1$ and $2$ pulsation modes for each model according to the adiabatic equations of non-radial stellar oscillations \citep{1989nos..book.....U}. For a comparison of the observed ($P_i^{\mathrm{obs}}$) and calculated ($P_i^{\mathrm{calc}}$) eigenmodes, we applied the \textsc{fitper} tool of \citet{2007PhDT........13K}, which calculates the root mean square ($\sigma_\mathrm{{rms}}$) for every model as follows:

\begin{equation}
\sigma_\mathrm{{rms}} = \sqrt{\frac{\sum_{i=1}^{N} (P_i^{\mathrm{calc}} - P_i^{\mathrm{obs}})^2}{N}},
\label{equ1}
\end{equation}

\noindent where \textit{N} is the number of observed periods; that is, $\sigma_\mathrm{{rms}}$ characterises the goodness of the fits.

First, we built a coarse (master) model grid covering a wide parameter space in the main physical properties: stellar mass and effective temperature. We scanned the parameter space as follows: $T_{\mathrm{eff}}$, $M_*$, $M_\mathrm{{env}}$ (the mass of the envelope, determined by the location of the base of the mixed helium and carbon layer), $M_\mathrm{H}$, $X_\mathrm{{He}}$ (the helium abundance in the C/He/H region), and $X_\mathrm{O}$ (the central oxygen abundance). While we scanned several parameters, we fixed the mass of the helium layer ($M_{\mathrm{He}}$) at $10^{-2}\,M_*$. This is the theoretical maximum for $M_{\mathrm{He}}$ and we did not vary it, considering the results of \citet{2012MNRAS.420.1462R}. These authors found that $M_{\mathrm{He}}$ can be lower by as much as a factor of 3-4  than the values based on to evolutionary calculations -- but not orders of magnitudes lower; therefore, it does not affect the periods substantially. Table~\ref{tabl:master} lists the parameter space we covered by the master grid, and the corresponding step sizes. We note that we calculated the possible pulsation periods of the models in the 70--1500\,s period range.

\begin{table}
\centering
\caption{Physical parameters varied while building the master grid. The step sizes applied are in parentheses.}
\label{tabl:master}
\begin{tabular}{lr}
\hline
\hline
$T_{\mathrm{eff}}$ [K] & $10\,000 - 14\,000$ [250]\\
$M_*$ [$M_{\odot}$] & $0.35 - 0.90$ [0.5]\\
-log$(M_\mathrm{{env}}/M_*)$ & $1.5 - 1.9$ [0.1]\\
-log$(M_{\mathrm{He}}/M_*)$ & $2$ [fixed]\\
-log$(M_\mathrm{H}/M_*)$ & $4 - 9$ [$1.0$]\\ 
$X_\mathrm{{He}}$ & $0.5 - 0.9$ [0.1]\\
$X_\mathrm{O}$ & $0.5 - 0.9$ [0.1]\\
\hline
\end{tabular}
\end{table}

In our investigation of this coarse grid, we considered six different cases with different restrictions on the degree of the pulsation modes: (a) we assumed that all 18 periods are $\ell=1$; (b) We treated the dominant periods of both detected by this work and listed in \citet{2006ApJ...640..956M} as $\ell=1$ modes, while we let the other modes to be either $\ell=1$ or $\ell=2$. In other words, in this case, both the 228.3 and 1058.4\,s modes were required to be $\ell=1$ periods during the fitting process; (c) We assumed that the 228.3\,s mode is an $\ell=1$, while we did not fix the $\ell$ values of the other modes; (d) Similarly to the previous case, we assumed that at least one mode should be $\ell=1$, but this time this mode was the 1058.4\,s period. (e) We selected models with at least ten $\ell=1$ modes, that is, we assumed that more than half of the modes are $\ell=1$. (f) Finally, we assumed only that all observed periods are either $\ell=1$ or $\ell=2$ modes, but we did not apply any further restriction on the degree of the modes.

Our results show the followings: (a) We did not find any solution with all detected periods being $\ell=1$ degree modes with better than $\sigma_\mathrm{{rms}} = 15\,$s; (b) With two or more $\ell=1$ modes, the best-fit model has the physical parameters of $T_{\mathrm{eff}} = 12\,250\,$K and $M_*=0.90\,M_{\odot}$ ($\sigma_\mathrm{{rms}} = 2.69\,$s); (c) Assuming the 228.3\,s period mode to be $\ell=1$ degree, the best fit model was found at $T_{\mathrm{eff}} = 13\,500\,$K and $M_*=0.80\,M_{\odot}$ ($\sigma_\mathrm{{rms}} = 2.52\,$s); (d) Assuming the 1058.4\,s period mode to be $\ell=1$ degree, the best-fit model was found at $T_{\mathrm{eff}} = 14\,000\,$K and $M_*=0.85\,M_{\odot}$ ($\sigma_\mathrm{{rms}} = 2.58\,$s); (e) The best-fit model that has at least 10 $\ell=1$ degree modes is $T_{\mathrm{eff}} = 13\,750\,$K and $M_*=0.90\,M_{\odot}$ ($\sigma_\mathrm{{rms}} = 2.70\,$s); (f) Without any restriction on the $\ell$ values, the best-fit model was found to be the same as in case (c). 

\begin{figure}
\centering
\includegraphics[width=0.45\textwidth]{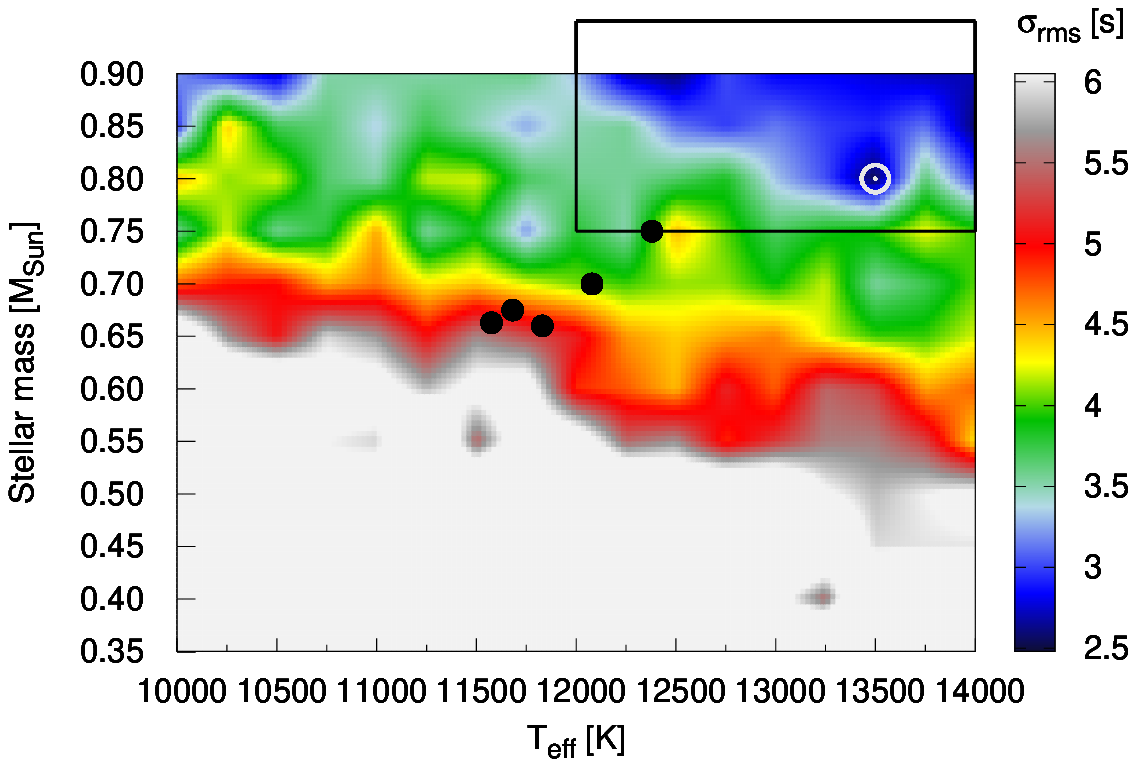}
\caption{Models on the $T_{\mathrm{eff}} - M_*$ plane of the coarse grid. We show the goodness of the fit of each grid-point with no restrictions on the degree of the observed pulsation modes. The model with the lowest $\sigma_\mathrm{{rms}}$ is denoted with a white open circle, while the spectroscopic solutions are marked with black dots. The parameter space further investigated by a higher resolution sub-grid is indicated by a black square.}{\label{fig:coarse_c}}
\end{figure}

In sum, our best fit solutions are situated at relative high effective temperatures, in the range of $\sim 12\,000 - 14\,000\,$K, and also show high stellar masses between $0.8$ and $0.9\,M_{\odot}$. That is, our fit results suggest \gd\ being a hot and massive star, hotter and more massive than it was found by previous spectroscopic measurements. The spectroscopic solutions obtained previously on \gd\ are listed in Table~\ref{tabl:spec}, while Fig.~\ref{fig:coarse_c} demonstrates the first stage of our search for the best fit model utilising a master grid.

\begin{table}
\centering
\caption{Spectroscopic values for \gd. The source of this list is the Montreal White Dwarf Database \citep{2017ASPC..509....3D}.}
\label{tabl:spec}
\begin{tabular}{lrr}
\hline
\hline
$T_{\mathrm{eff}}$ [K] & $M_*$ [$M_{\odot}$] & Ref.\\
\hline
$11\,830 \pm 174$ & $0.66 \pm 0.03$ & \citet{2005ApJS..156...47L} \\
$12\,380 \pm 190$ & $0.75 \pm 0.03$ & \citet{2011ApJ...743..138G} \\
$12\,080 \pm 179$ & $0.70 \pm 0.03$ & \citet{2015ApJS..219...19L} \\
$11\,681 \pm 72$ & $0.68 \pm 0.01$ & \citet{2020MNRAS.499.1890M} \\
$11\,575 \pm 038$ & $0.663 \pm 0.004$ & \citet{2020ApJ...898...84K} \\
\hline
\end{tabular}
\end{table}

Considering these results, we built a second, higher-resolution subgrid to further investigate the possible asteroseismic solutions for \gd. Table~\ref{tabl:refined} lists the parameter space we covered by this subgrid and the corresponding step sizes.  

\begin{table}
\centering
\caption{Physical parameters varied while building a refined grid. The step sizes applied are in parentheses.}
\label{tabl:refined}
\begin{tabular}{lr}
\hline
\hline
$T_{\mathrm{eff}}$ [K] & $12\,000 - 14\,000$ [100]\\
$M_*$ [$M_{\odot}$] & $0.75 - 0.95$ [0.1]\\
-log$(M_\mathrm{{env}}/M_*)$ & $1.5 - 1.9$ [0.1]\\
-log$(M_{\mathrm{He}}/M_*)$ & $2$ [fixed]\\
-log$(M_\mathrm{H}/M_*)$ & $4 - 9$ [$0.5$]\\ 
$X_\mathrm{{He}}$ & $0.5 - 0.9$ [0.1]\\
$X_\mathrm{O}$ & $0.5 - 0.9$ [0.1]\\
\hline
\end{tabular}
\end{table}

Utilising the higher-resolution subgrid, we found solutions for all six cases. We list the physical parameters of the best fit models both for the master grid and the sub-grid in Table~\ref{tabl:params}.

We can check the reliability of our best-fit solutions listed in Table~\ref{tabl:params} considering the known geometric distance of the star provided by the Gaia space telescope. For our target, this geometric distance is $d_{Gaia} = 33.939^{+0.027}_{-0.030}$\,pc \citep{2021AJ....161..147B}, which is based on the Gaia early third release EDR3 \citep{2021A&A...649A...1G}. The method we followed to determine the seismic distances of the selected models utilising the luminosity of the given model and the apparent visual magnitude of the star is detailed, for instance, in \citet{2019A&A...632A..42B} and \citet{2021A&A...651A..14B}. For the apparent visual magnitude of \gd, the fourth US Naval Observatory CCD Astrograph Catalog \citep{2012yCat.1322....0Z} gives $m_{\mathrm V} = 14.584\pm0.06\,$mag. We listed the seismic distances derived for the best-fit models for the different cases in the last column of Table~\ref{tabl:params}.

We found that the best matching model to the Gaia geometric distance is the $T_{\mathrm{eff}} = 13\,500\,$K and $M_*=0.80\,M_{\odot}$, with the seismic distance of 34.19$\pm$0.94\,pc. There is another model with $T_{\mathrm{eff}} = 14\,000\,$K, $M_*=0.85\,M_{\odot}$, and the seismic distance of 33.25$\pm$0.92\,pc, in which case the seismic and Gaia geometric distances are practically the same within the errors. In the other cases listed in Table~\ref{tabl:params}, the distances suggest a star closer to us than we found by the Gaia data.

Being somewhat closer in effective temperature and stellar mass to the spectroscopic values, we selected the $T_{\mathrm{eff}} = 13\,500\,$K and $M_*=0.80\,M_{\odot}$ model for further investigation. Table~\ref{tabl:periods} summarises the calculated and observed periods for this model.

\begin{table*}
\centering
\caption{Physical parameters and seismic distances ($d$) of the best-fit models. Short summary on the different cases: case (a) -- all modes are $\ell=1$, case (b) -- the 228.3 and 1058.4\,s modes are $\ell=1$, case (c) -- the 228.3\,s mode is $\ell=1$, case (d) -- the 1058.4\,s mode is $\ell=1$, case (e) -- at least ten $\ell=1$ modes, case (f) -- no restrictions on the modes' $\ell$ values.}
\label{tabl:params}
\begin{tabular}{rrrrrrrrr}
\hline
\hline
\multicolumn{1}{c}{$T_{\mathrm{eff}}$ [K]} & \multicolumn{1}{c}{$M_*$ [$M_{\odot}$]} & \multicolumn{1}{c}{-log$(M_\mathrm{{env}}/M_*)$} & \multicolumn{1}{c}{-log$(M_\mathrm{{He}}/M_*)$} & \multicolumn{1}{c}{-log$(M_\mathrm{{H}}/M_*)$} & \multicolumn{1}{c}{$X_\mathrm{{He}}$} & \multicolumn{1}{c}{$X_\mathrm{{O}}$} & \multicolumn{1}{c}{$\sigma_\mathrm{{rms}}$ [s])} & \multicolumn{1}{c}{$d$ [pc]} \\
\hline
\multicolumn{9}{l}{master grid -- case (b):} \\
12\,250 & 0.90 & 1.8 & 2.0 & 4.0 & 0.9 & 0.8 & 2.69 & 27.96$\pm$0.77 \\
\multicolumn{9}{l}{master grid -- case (c) -- same as case (f):} \\
13\,500 & 0.80 & 1.6 & 2.0 & 4.0 & 0.8 & 0.6 & 2.52 & 34.19$\pm$0.94 \\
\multicolumn{9}{l}{master grid -- case (d):} \\
14\,000 & 0.85 & 1.6 & 2.0 & 4.0 & 0.8 & 0.6 & 2.58 & 33.25$\pm$0.92 \\
\multicolumn{9}{l}{master grid -- case (e):} \\
13\,750 & 0.90 & 1.5 & 2.0 & 4.0 & 0.5 & 0.8 & 2.70 & 30.72$\pm$0.85 \\
\multicolumn{9}{l}{subgrid -- case (a):} \\
14\,000 & 0.94 & 1.6 & 2.0 & 4.5 & 0.9 & 0.5 & 5.76 & 29.45$\pm$0.81 \\
\multicolumn{9}{l}{subgrid -- case (b):} \\
13\,000 & 0.94 & 1.6 & 2.0 & 4.0 & 0.7 & 0.9 & 2.04 & 27.81$\pm$0.77 \\
\multicolumn{9}{l}{subgrid -- case (c):} \\
12\,200 & 0.91 & 1.9 & 2.0 & 4.0 & 0.9 & 0.6 & 2.03 & 27.52$\pm$0.76 \\
\multicolumn{9}{l}{subgrid -- case (d) -- sames as case (f):} \\
12\,400 & 0.93 & 1.7 & 2.0 & 4.0 & 0.8 & 0.6 & 1.83 & 27.21$\pm$0.75 \\
\multicolumn{9}{l}{subgrid -- case (e):} \\
12\,400 & 0.93 & 1.7 & 2.0 & 4.0 & 0.9 & 0.6 & 2.13 & 27.21$\pm$0.75 \\[1mm]
 & & & & & & & Gaia: & 33.939$^{+0.027}_{-0.030}$ \\[1mm]
\hline
\end{tabular}
\end{table*}

\begin{table}
\centering
\caption{Observed and calculated periods of the selected model (see master grid -- case (c) in Table~\ref{tabl:params}). 
In the last column, we also list the period differences.}
\label{tabl:periods}
\begin{tabular}{rrcc}
\hline
\hline
\multicolumn{1}{c}{Obs. period [s]} & \multicolumn{1}{c}{Calc. period [s]} & $\ell$ & \multicolumn{1}{c}{Difference [s]} \\
\hline
105.2 & 109.0& 2 & 3.8 \\
222.6 & 218.9& 2 & 3.7 \\
228.3 & 227.9& 1 & 0.4 \\
286.0 & 287.5& 2 & 1.5 \\
299.6 & 301.7& 1 & 2.1 \\
346.0 & 342.4& 1 & 3.6 \\
593.3 & 594.3& 2 & 1.1 \\
633.1 & 632.8& 2 & 0.3 \\
853.2 & 856.3& 2 & 3.1 \\
871.5 & 873.7& 1 & 2.3 \\
924.7 & 925.9& 2 & 1.2 \\
957.9 & 954.5& 1 & 3.4 \\
976.0 & 974.1& 2 & 1.9 \\
1007.0 & 1007.2& 2 & 0.2 \\
1058.4 & 1060.3& 2 & 1.9 \\
1069.9 & 1066.3& 1 & 3.6 \\
1088.0 & 1086.8& 2 & 1.2 \\
1151.0 & 1146.9& 2 & 4.1 \\
\hline
\end{tabular}
\end{table}

\begin{figure}
\centering
\includegraphics[width=0.45\textwidth]{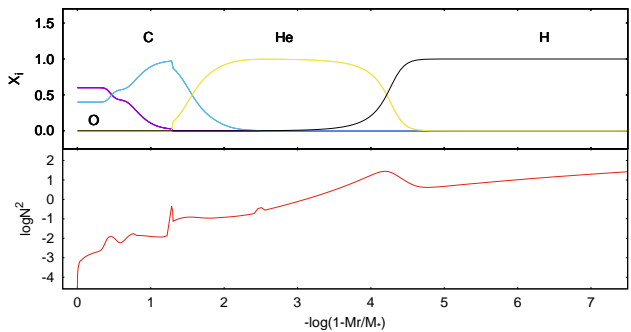}
\caption{Chemical composition profiles in fractional abundances, and the corresponding logarithm of the squared Brunt-V\"ais\"al\"a frequency ($\mathrm{log}\,N^2$) for our selected model by the refined grid ($T_{\mathrm{eff}}=13\,500\,$K, $M_*=0.80\,M_{\odot}$, $-$log$M_\mathrm{{env}}=1.6$, $-$log$M_\mathrm{He}=2.0$, $-$log$M_\mathrm{H}=4.0$, $X_\mathrm{{He}}=0.8$, $X_\mathrm{O}=0.6$).}{\label{fig:profile}}
\end{figure}

We plot the chemical composition profiles and also the Brunt--V\"ais\"al\"a frequency profile of our accepted model in Fig.~\ref{fig:profile}. As Table~\ref{tabl:params} shows, we obtained 60 per cent oxygen abundance for the core. Comparing it with the expected $\sim 65\%$  value by the evolutionary calculations of \citet{2012MNRAS.420.1462R}, our solution is close to this probable core oxygen abundance.

We note that considering our model solutions, we obtain models in which the chemical profiles show a pure carbon buffer, in spite of the fact that fully evolutionary models are not in agreement with this characteristics. Instead, the pure carbon buffer can disappear by chemical diffusion (see e.g., \citealt{2019A&A...630A.100D}).

\section{Summary and discussion}
\label{sect:sum}

The long-known \gd\ is one of the brightest ZZ~Ceti type pulsating white dwarf, however, no detailed investigation on its pulsational behaviour and asteroseismology have been published thus far. That is the reason why we selected \gd\ for a long-time ground-based follow-up measurement, resulting in 30 nights of observations and spanning more than 119 observing hours. Our extended observations enabled the detection of 11 eigenmodes for seismology and we also revealed that the peaks detected in the TESS light curve are Nyquist aliases of the frequencies derived based on the ground-based observations. This highlights the usefulness of ground-based observations, even when space-based measurements are available. Moreover, with the modes detected by our ground-based measurements, and given that we can complete our set of modes with the ones presented by \citet{2006ApJ...640..956M}, \gd\ joined the small group of known white dwarf pulsators that are rich in oscillation modes.

We performed a preliminary asteroseismic analysis of \gd\ by comparing the observed and calculated pulsating periods. Our preferred model is the one with $T_{\mathrm{eff}}=13\,500\,$K and $M_*=0.80\,M_{\odot}$, that is to say, according to our seismic fittings, \gd\ may be at the blue edge of the empirical ZZ~Ceti instability strip, (see e.g. Fig.~3 in \citealt{2017ApJS..232...23H}). 

Considering the frequency separation ($\delta f$) of the well-resolved doublet found by the TESS light curve, we can estimate the rotation period of the star. This is because, in the case of slow rotation, the frequency differences of the
$m = -1, 0, 1$ rotationally split components can be approximated to first order by the following relation:

\begin{equation}
\delta f_{k,l,m} = \delta m (1 - C_{k,l}) \Omega,
\label{equ2}
,\end{equation}

\noindent where the coefficient $C_{k,l} \approx 1/\ell(\ell +1)$ for high-overtone ($k \gg l$) $g$-modes and $\Omega$ is the (uniform) rotation frequency. The frequency separation of the TESS doublet of the  \gd\ data is $11.81\,\mu$Hz, while the corresponding $C_{k,l}$ value is $0.44$, as these are low-radial-order frequencies. Finally, we obtain a $P = 0.55\,\mathrm{d} = 13.17\,$h rotation period for \gd. Thanks to the TESS observations, this is the first time we were able to derive the rotational rate for this star.


\begin{acknowledgements}

The authors thank the anonymous referee for the constructive comments and recommendations on the manuscript.

The authors also thank the efforts of Zsolt Reg\'aly (Konkoly Observatory) considering the development of the observing tools for the EMCCD camera.

The authors acknowledge the financial support of the Lend\"ulet Program of the Hungarian Academy of Sciences, projects No. LP2018-7/2022, and LP2012-31. This research was supported by the KKP-137523 `SeismoLab' \'Elvonal grant of the Hungarian Research, Development and Innovation Office (NKFIH).

The authors are also grateful for the usage of ELKH Cloud (see \citealt{H_der_2022}; \url{https://science-cloud.hu/}) which helped us achieve the results published in this paper.

ZsB acknowledges the support by the J\'anos Bolyai Research Scholarship of the Hungarian Academy of Sciences.

This paper includes data collected with the TESS mission, obtained from the MAST data archive at the Space Telescope Science Institute (STScI). Funding for the TESS mission is provided by the NASA Explorer Program. STScI is operated by the Association of Universities for Research in Astronomy, Inc., under NASA contract NAS 5–26555.

This work has made use of data from the European Space Agency (ESA) mission {\it Gaia} (\url{https://www.cosmos.esa.int/gaia}), processed by the {\it Gaia} Data Processing and Analysis Consortium (DPAC, \url{https://www.cosmos.esa.int/web/gaia/dpac/consortium}). Funding for the DPAC
has been provided by national institutions, in particular the institutions participating in the {\it Gaia} Multilateral Agreement.
\end{acknowledgements}



\bibliographystyle{aa} 
\bibliography{gd99} 

\end{document}